\UseRawInputEncoding 
\documentclass[twocolumn]{aastex62}

\newcommand{\lsim}{\raisebox{-0.13cm}{~\shortstack{$<$ \\[-0.07cm] $\sim$}}~}
\newcommand{\gsim}{\raisebox{-0.13cm}{~\shortstack{$>$ \\[-0.07cm] $\sim$}}~}
\newcommand{\hbo}{$\rm (H\beta + [OIII])$~}

\received{...}
\revised{...}
\accepted{...}
\submitjournal{ApJ}

\shorttitle{The Relation between Strong \hbo Emission, Star Formation and Burstiness}
\shortauthors{Caputi et al.}
\begin{document}

\title{MIDIS: The Relation between Strong \hbo Emission, Star Formation and Burstiness Around the Epoch of Reionization}

\correspondingauthor{K. I. Caputi}
\email{karina@astro.rug.nl}

\author[0000-0001-8183-1460]{K. I. Caputi}
\affil{Kapteyn Astronomical Institute, University of Groningen, P.O. Box 800, 9700AV Groningen, The Netherlands}
\affil{Cosmic Dawn Center (DAWN), Copenhagen, Denmark}

\author[0000-0002-5104-8245]{P. Rinaldi}
\affil{Kapteyn Astronomical Institute, University of Groningen, P.O. Box 800, 9700AV Groningen, The Netherlands}

\author[0000-0001-8386-3546]{E. Iani}
\affil{Kapteyn Astronomical Institute, University of Groningen, P.O. Box 800, 9700AV Groningen, The Netherlands}

\author[0000-0003-4528-5639]{P. G. P\'erez-Gonz\'alez}
\affiliation{Centro de Astrobiolog\'{\i}a (CAB), CSIC-INTA, Ctra. de Ajalvir km 4, Torrej\'on de Ardoz, E-28850, Madrid, Spain}

\author[0000-0002-3005-1349]{G. \"Ostlin}
\affiliation{Department of Astronomy, Stockholm University, Oscar Klein Centre, AlbaNova University Centre, 106 91 Stockholm, Sweden}

\author[0000-0002-9090-4227]{L. Colina}
\affiliation{Centro de Astrobiolog\'{\i}a (CAB), CSIC-INTA, Ctra. de Ajalvir km 4, Torrej\'on de Ardoz, E-28850, Madrid, Spain}
\affil{Cosmic Dawn Center (DAWN), Copenhagen, Denmark}

\author[0000-0002-2554-1837]{T. R. Greve}
\affil{Cosmic Dawn Center (DAWN), Copenhagen, Denmark}
\affiliation{DTU-Space, Elektrovej, Building 328 , 2800, Kgs. Lyngby, Denmark}

\author[0000-0000-0000-0000]{H. U. N\o{}rgaard-Nielsen}
\affil{Cosmic Dawn Center (DAWN), Copenhagen, Denmark}
\affiliation{DTU-Space, Elektrovej, Building 328 , 2800, Kgs. Lyngby, Denmark}

\author[0000-0000-0000-0000]{G. S. Wright}
\affiliation{UK Astronomy Technology Centre, Royal Observatory Edinburgh,
Blackford Hill, Edinburgh EH9 3HJ, UK}

\author[0000-0002-7093-1877]{J. \'Alvarez-M\'arquez}
\affiliation{Centro de Astrobiolog\'{\i}a (CAB), CSIC-INTA, Ctra. de Ajalvir km 4, Torrej\'on de Ardoz, E-28850, Madrid, Spain}

\author[0000-0000-0000-0000]{A. Eckart}
\affiliation{I.Physikalisches Institut der Universit\"at zu K\"oln, Z\"ulpicher Str. 77,
50937 K\"oln, Germany}

\author[0000-0002-4571-2306]{J. Hjorth}
\affiliation{DARK, Niels Bohr Institute, University of Copenhagen, Jagtvej 128,
2200 Copenhagen, Denmark}

\author[0000-0002-0690-8824]{A. Labiano}
\affiliation{Telespazio UK for the European Space Agency (ESA), ESAC, Camino Bajo del Castillo s/n, 28692 Villanueva de la Ca\~nada, Spain}

\author[0000-0000-0000-0000]{O. Le F\`evre}
\affiliation{Aix Marseille Universit\'e, CNRS, LAM (Laboratoire d'Astrophysique de Marseille) UMR 7326, 13388, Marseille, France}

\author[0000-0003-4793-7880]{F. Walter}
\affiliation{Max-Planck-Institut f\"ur Astronomie, K\"onigstuhl 17, 69117 Heidelberg, Germany}

\author[00000-0001-5434-5942]{P. van der Werf}
\affiliation{Leiden Observatory, Leiden University, PO Box 9513, 2300 RA Leiden, The Netherlands}

\author[0000-0002-3952-8588]{L. Boogaard}
\affiliation{Max-Planck-Institut f\"ur Astronomie, K\"onigstuhl 17, 69117 Heidelberg, Germany}

\author[0000-0001-6820-0015]{L. Costantin}
\affiliation{Centro de Astrobiolog\'{\i}a (CAB), CSIC-INTA, Ctra. de Ajalvir km 4, Torrej\'on de Ardoz, E-28850, Madrid, Spain}

\author[0000-0003-2119-277X]{A. Crespo G\'{o}mez}
\affiliation{Centro de Astrobiolog\'{\i}a (CAB), CSIC-INTA, Ctra. de Ajalvir km 4, Torrej\'on de Ardoz, E-28850, Madrid, Spain}

\author[0000-0001-9885-4589]{S. Gillman}
\affil{Cosmic Dawn Center (DAWN), Copenhagen, Denmark}
\affiliation{DTU-Space, Elektrovej, Building 328 , 2800, Kgs. Lyngby, Denmark}

\author[0000-0002-2624-1641]{I. Jermann}
\affil{Cosmic Dawn Center (DAWN), Copenhagen, Denmark}
\affiliation{DTU-Space, Elektrovej, Building 328 , 2800, Kgs. Lyngby, Denmark}

\author[0000-0001-5710-8395]{D. Langeroodi}
\affiliation{DARK, Niels Bohr Institute, University of Copenhagen, Jagtvej 128, 2200 Copenhagen, Denmark}

\author[0000-0003-0470-8754]{J. Melinder}
\affiliation{Department of Astronomy, Stockholm University, Oscar Klein Centre, AlbaNova University Centre, 106 91 Stockholm, Sweden}

\author[0000-0000-0000-0000]{F. Peissker}
\affiliation{I.Physikalisches Institut der Universit\"at zu K\"oln, Z\"ulpicher Str. 77,
50937 K\"oln, Germany}

\author[0000-0001-9818-0588]{M. G\"udel}
\affiliation{Dept. of Astrophysics, University of Vienna, T\"urkenschanzstr. 17, A-1180 Vienna, Austria}
\affiliation{Max-Planck-Institut f\"ur Astronomie, K\"onigstuhl 17, 69117 Heidelberg, Germany}
\affiliation{ETH Z\"urich, Institute for Particle Physics and Astrophysics, Wolfgang-Pauli-Str. 27, 8093 Z\"urich, Switzerland}

\author[0000-0002-1493-300X]{Th.~Henning}
\affiliation{Max-Planck-Institut f\"ur Astronomie, K\"onigstuhl 17, 69117 Heidelberg, Germany}

\author[0000-0000-0000-0000]{P.~O.~Lagage}
\affiliation{AIM, CEA, CNRS, Universit\'e Paris-Saclay, Universit\'e Paris Diderot, Sorbonne Paris Cit\'e, F-91191 Gif-sur-Yvette, France}

\author[0000-0002-2110-1068]{T.~P.~Ray}
\affiliation{Dublin Institute for Advanced Studies, 31 Fitzwilliam Place, D02 XF86 Dublin, Ireland}

%\author{}
%\affil{}

%% Note that the \and command from previous versions of AASTeX is now
%% depreciated in this version as it is no longer necessary. AASTeX 
%% automatically takes care of all commas and "and"s between authors names.

%% AASTeX 6.1 has the new \collaboration and \nocollaboration commands to
%% provide the collaboration status of a group of authors. These commands 
%% can be used either before or after the list of corresponding authors. The
%% argument for \collaboration is the collaboration identifier. Authors are
%% encouraged to surround collaboration identifiers with ()s. The 
%% \nocollaboration command takes no argument and exists to indicate that
%% the nearby authors are not part of surrounding collaborations.

%% Mark off the abstract in the ``abstract'' environment. 
\begin{abstract}

We investigate the properties of strong \hbo emitters before and after the end of the Epoch of Reionization from $z=8$ to $z=5.5$. We make use of ultra-deep \textit{JWST/}NIRCam imaging in the Parallel Field of the MIRI Deep Imaging Survey (MIDIS) in the Hubble eXtreme Deep Field (P2-XDF), in order to select prominent \hbo emitters (with rest $\rm EW_0 \gsim 100 \, \AA$) at $z=5.5-7$, based on their flux density enhancement in the F356W band with respect to the spectral energy distribution continuum. We complement our selection with other  \hbo emitters from the literature at similar and higher ($z=7-8$) redshifts. We find (non-independent) anti-correlations between $\rm EW_0$\hbo and both galaxy stellar mass and age, in agreement with previous studies, and a positive correlation with specific star formation rate (sSFR). On the SFR-M$^\star$ plane, the \hbo emitters populate both the star-formation main sequence and the starburst region, which become indistinguishable at low stellar masses ($\rm log_{10}(M^\star) < 7.5$). We find tentative evidence for a non-monotonic relation between $\rm EW_0$\hbo  and SFR, such that both parameters correlate with each other at $\rm SFR \gsim 1 \, M_\odot / yr$, while the correlation flattens out at lower SFRs. This suggests that low metallicities producing high $\rm EW_0$\hbo could be important at low SFR values.  Interestingly, the properties of the strong emitters and other galaxies (33\% and 67\% of our $z=5.5-7$ sample, respectively) are similar, including, in many cases, high sSFR. Therefore, it is crucial to consider both emitters and non-emitters to obtain a complete picture of the cosmic star formation activity around the Epoch of Reionization.

 \end{abstract}

%% Keywords should appear after the \end{abstract} command. 
%% See the online documentation for the full list of available subject
%% keywords and the rules for their use.

%% From the front matter, we move on to the body of the paper.
%% Sections are demarcated by \section and \subsection, respectively.
%% Observe the use of the LaTeX \label
%% command after the \subsection to give a symbolic KEY to the
%% subsection for cross-referencing in a \ref command.
%% You can use LaTeX's \ref and \label commands to keep track of
%% cross-references to sections, equations, tables, and figures.
%% That way, if you change the order of any elements, LaTeX will
%% automatically renumber them.

%% We recommend that authors also use the natbib \citep
%% and \citet commands to identify citations.  The citations are
%% tied to the reference list via symbolic KEYs. The KEY corresponds
%% to the KEY in the \bibitem in the reference list below. 

\section{Introduction} \label{sec:intro}

Investigating the properties of galaxies in the early Universe is necessary to understand the first steps of galaxy evolution and their link to the process of Reionization. Until very recently, these studies were limited to the brightest galaxies at rest-frame UV wavelengths, given the lack of sensitive telescopes operating at $\lambda \gsim 2 \, \rm \mu m$. The new \textit{JWST} observations are now radically transforming this field by giving us access to much fainter sources, including the precursor seed units that have eventually grown into more massive galaxies at later cosmic times. 

Two of the most fundamental galaxy properties that define galaxy growth are the already assembled stellar mass ($\rm M_\star$) and the ongoing star formation rate (SFR). Although these two properties are known to be related \citep[e.g., ][]{Brinchmann_2004, Speagle_2014, RenziniPeng_2015}, it is unclear whether the corresponding physical conditions for star formation are the same at all scales, particularly at high redshifts. Investigating the relation between different galaxy physical parameters and these more fundamental properties is crucial to explain how  galaxy evolution took place at early cosmic times.

The search for line emitters provides a shortcut for selecting star-forming galaxies. Especially the presence of the brightest emission lines, such as the Balmer lines, as well as [OII]3727 and [OIII]4959, 5007, helps boosting the galaxy detectability. However, the detection of such emission lines appearing in the rest-frame optical regime has traditionally been difficult beyond intermediate redshifts, also because of the wavelength coverage and sensitivity of existing telescopes. Now \textit{JWST} has turned the study of rest-optical emission lines to be routinely possible in galaxies up to very high redshifts.

Many studies conducted over the past decade concluded that some emission lines become increasingly important with redshift, i.e., they are more luminous and have higher equivalent widths (EW) \citep[e.g., ][]{RobertsBorsani_2016, debarros_2019, Matthee_2023}. This is particularly the case of the \hbo line complex \citep[e.g.][]{Smit_2014, Khostovan_2016, Reddy_2018, Endsley_2021}. Most of these studies have been based on photometric data, as a less costly alternative to spectroscopy. The presence of prominent  emission lines (i.e., emission lines with high equivalent width)  can be inferred from photometric measurements via the flux density excess with respect to the spectral continuum, which is produced in the photometric band in which the line is observed \citep[e.g.,][]{Sobral_2013, Smit_2014, Smit_2016, Caputi_2017}.

The success of this technique has now triggered a number of studies of high-redshift line emitters based on \textit{JWST} images, reaching galaxies up to the Epoch of Reionization ($z \sim 7$). These works have analysed the dependence of the line emission on general galaxy properties, such as rest-UV absolute magnitudes and stellar mass, and inferred the ionizing photon production efficiency to constrain the role of the emitters in the process of Reionization \citep[e.g., ][]{Prieto_Lyon_2023,Endsley_2023a, Rinaldi_2023}. Yet, it has been recently pointed out that line emitters may provide a biased view of the star-formation activity at high redshifts \citep{Sun_2023}. Putting them in the context of \textit{all} galaxies present at the same redshifts could, thus, be necessary to understand their importance and achieve a complete picture of early galaxy evolution.

Another aspect that has become evident in the past years is the increasing importance of starburst galaxies with redshift \citep[e.g.][]{Caputi_2017, Rinaldi_2022}. These are galaxies whose star formation activity is temporarily enhanced, such that they are placed significantly above the so-called main sequence (MS) of star formation \citep[e.g.,][]{Peng_2010, Speagle_2014, Salmon_2015, Rinaldi_2022}. Although in the literature there are different definitions for starburst galaxies, a clear way to select them is via their specific star formation rates (sSFR), which has been empirically defined to be $\rm log_{10}(sSFR(yr^{-1}))>-7.6$ \citep{Caputi_2017, Caputi_2021}. This implies a stellar-mass doubling time (i.e., the inverse of the sSFR) of $\lsim 4 \times 10^7 \, \rm yr$, which is roughly compliant with the timescales for starburst episodes studied in the local Universe \citep[e.g.,][]{Heckman_1998, Kennicutt_1998, Leitherer_2002, Ostlin_2003}. The incidence of starbursts is higher amongst low stellar-mass galaxies \citep{Bisigello_2018} and the recent JWST studies provide hints that the starburst mode of star formation could be very important in the early Universe \citep[e.g.,][]{Dressler_2023, Endsley_2023b}.

The goal of this paper is to investigate the connection between prominent \hbo emission and other properties, including the SFR and sSFR, in galaxies before and after the end of the Epoch of Reionization from $z=8$ to $z=5.5$. For this purpose we analyse the deepest existing JWST imaging data, which allow us to select galaxies down to unprecedented low stellar mass limits at those redshifts. Moreover, we compare the properties of these emitters to those of all other galaxies at the same redshifts, in order to understand their role in the early steps of galaxy evolution. Throughout this paper we adopt a cosmology  with $\rm H_0=70 \,{\rm km \, s^{-1} Mpc^{-1}}$, $\rm \Omega_M=0.3$ and $\rm \Omega_\Lambda=0.7$. All magnitudes in this paper are total and are expressed in the AB system \citep{Oke_1983}. Stellar masses and SFRs refer to a \citet{Chabrier_2003} initial mass function.

\section{Datasets} \label{sec:data}

\begin{figure}[t]
\center{
\includegraphics[width=1.1\linewidth, keepaspectratio]{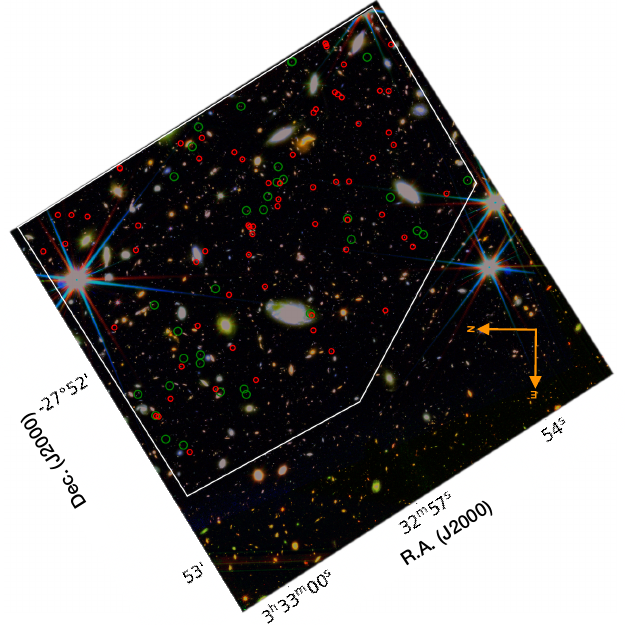}
\caption{RGB composite image of the P2 field with NIRCam coverage. The region delimited with a white line has the deepest coverage in the NIRCam bands and is the field considered in this work. Green circles indicate the location of our \hbo emitters at $z=5.5-7.0$, while the red circles correspond to all other sources in the same redshift range.} \label{fig:field}
}
\end{figure}

We made use of the ultra-deep \textit{JWST/}NIRCam images that have been taken in parallel with the \textit{JWST} Guaranteed Time Observations (GTO) program {\it MIRI Deep Imaging Survey} (MIDIS; PID: 1283, PI: G\"oran \"Ostlin) in the Hubble eXtreme Deep Field (H-XDF). We also analysed data from the Next Generation Deep Extragalactic Exploratory Public (NGDEEP; PID: 2079; PIs: S.~Finkelstein, C.~Papovich and N.~Pirzkal) survey. All these NIRCam images partly or entirely cover the second HUDF parallel field \citep[hereafter P2;][]{Whitaker_2019}. Observations have been taken in a total of six \textit{JWST/}NIRCam broadbands: F115W, F150W, F277M, and F356M (MIDIS), and F200W and F444W (NGDEEP). More information about these data can be found in \citet{PerezGonzalez_2023} and \citet{Austin_2023}. Here we restrict our analysis to the $\simeq$3.3~arcmin$^2$ area that has maximum homogeneously deep coverage in the NIRCam filters  (P2/NIRCam hereafter; Fig.~\ref{fig:field}).

We have processed all these NIRCam images with a modified version of the official \textit{JWST} pipeline\footnote{The official \textit{JWST} pipeline is available  \href{https://github.com/spacetelescope/jwst}{here}} (based on \texttt{jwst 1.8.2} and Calibration Reference Data System pipeline mapping (CRDS; pmap) \texttt{1084})). Detailed information about the reference files is available at \href{https://jwst-crds.stsci.edu}{STScI/CRDS}. Compared to the official \textit{JWST} pipeline, our version includes a number of extra steps to deal with some of the problems that still affected the official software. We minimized the impact of the so-called `snowballs' and 'wisps'\footnote{For more information see \href{https://jwst-docs.stsci.edu/\#gsc.tab=0}{\textit{JWST's} documentation webpage}}, as well as the \textit{1/f} noise and residual cosmic rays. After reducing all the NIRCam images, we drizzled and mosaiced all the resulting calibrated files to 0.03\arcsec/pixel, which is the final pixel scale of our images for science analysis. All these final images have been aligned to the Hubble Legacy Fields (HLF) catalogue\footnote{The HLF catalogue is available  \href{https://archive.stsci.edu/hlsps/hlf}{here}}.

We tested our data reduction by comparing the photometry for the brightest sources ($<$ 24 mag) in all the NIRCam filters, following the same approach that we already adopted in \citet{Rinaldi_2023}. To do that, we produced two versions of the NIRCam images, with and without the aforementioned extra steps. Finally, we extracted the sources by using the software \texttt{Source Extractor} \citep[SExtractor,][]{SExtractor} and compared their photometry. This test confirmed that our extra steps do not introduce any systematic effect in the photometry.

As a complement, we also considered  \textit{Hubble Space Telescope (HST)} images over the HUDF/P2 from the Hubble Legacy Field GOODS-S (HLF-GOODS-S)\footnote{The \textit{HST} images (0.03\arcsec/pixel) have been downloaded from the \href{https://archive.stsci.edu/prepds/hlf}{Space Telescope Science Institute's archive}.}. The HLF-GOODS-S data in HUDF/P2 is comprised of images in 10 \textit{HST} broad bands covering the optical (ACS/WFC F435W, F606W, F775W, F814W and F850LP), and near-infrared (WFC3/IR F098M, F105W, F125W, F140W and F160W). See \citet{Whitaker_2019} for more detailed information on these observations. In this work we only make use of the ACS/WFC images because of two reasons:  the WFC3/IR coverage in HUDF/P2 is not as homogeneous as for the ACS/WFC filters; and the WFC3/IR images are significantly shallower than the \textit{JWST/}NIRCam images. In summary, our finally considered image set has coverage in the wavelength range  0.4~$\mu$m through 4.4~$\mu$m,  with a total of 11 broad bands (5 \textit{HST} + 6 \textit{JWST} filters).

 Our JWST NIRCam imaging in HUDF/P2 is to our knowledge the deepest NIRCam data currently available: their depth is 30.2~mag and 30.8~mag ($5\sigma$), in F150W and F2777W, respectively \citep{PerezGonzalez_2023}.  This is about two magnitdes deeper than the NIRCam images utilised in \citet{Endsley_2023a} and about one magnitude deeper than the JEMS images \citep{Williams_2023}.

\section{Photometry and SED Fitting}

We performed the source detection and photometric measurements using the software \texttt{SExtractor} \citep{SExtractor}, which we used in dual mode on all bands. In all cases, for the source detection we adopted a super-stack image that we created by combining all NIRCam bands. To construct our photometric catalogue we used a combination of aperture photometry on 0.5"-diameter circular apertures and Kron apertures \citep[i.e., \texttt{MAG\_AUTO},][]{Kron_1980}, following a similar prescription to that adopted by \citet{Rinaldi_2022, Rinaldi_2023}. For sources with $\rm mag < 27$ , we chose the brightest amongst the circular-aperture flux (+ aperture correction) and the Kron flux. For fainter sources, we always adopted the circular-aperture flux (+ aperture correction).  We determined the limiting magnitude above which only aperture fluxes are considered based on tests performed on the \textit{HST} photometry (see \citet{Rinaldi_2023} for details). Finally, all our fluxes have been corrected for Galactic extinction.

We adopted a minimum error of 0.05~mag for all the HST photometry because \texttt{SExtractor} typically underestimates the photometric errors \citep[e.g.,][]{Sonnett_2013}. We decided to adopt this minimum error for the NIRCam images as well to account for possible uncertainties in the flux calibration.
For non-detected sources in any given band,  we estimated flux upper limits by performing empty-aperture statistics. We placed multiple (0.5\arcsec-diameter) random circular apertures on the corresponding background image to estimate the background r.m.s. (1$\sigma$),  which in our case is about 32.0-32.5~mag, depending on the NIRCam band.

We performed the SED fitting of our sources using the code \texttt{LePHARE} \citep{LePhare_2011}, following a similar prescription to that described in \citet{Rinaldi_2022, Rinaldi_2023}.  We considered the synthetic model templates by \citet[][hereafter BC03]{BC_2003}, making use of two different star formation histories (SFHs): a standard exponentially declining SFH (with 8 different $\tau$ values) and  a single stellar population (SSP). We adopted two distinct metallicity values, a solar metallicity (Z$_{\odot}$ = 0.02) and a fifth of solar metallicity (Z = 0.2Z$_{\odot}$ = 0.004). In addition, to take into account the strong contribution from nebular emission, we also considered \texttt{STARBURST99} templates \citep[][hereafter SB99]{SB99} for young galaxies (age $\leq 10^7 \, \rm yr$) with constant star formation histories.

We adopted the \citet{Calzetti_2001} reddening law in combination with the \citet{Leitherer_2002} prescription below 912 {\AA} to convolve the model templates and account for dust extinction. We used a colour-excess grid  $0 \leq E(B-V)\leq 1.5$, with a step of 0.1. We refer the reader to \citet{Rinaldi_2022, Rinaldi_2023} for further details on the SED fitting procedure.

\begin{figure}[h!]
\center{
\includegraphics[width=1.0\linewidth, keepaspectratio]{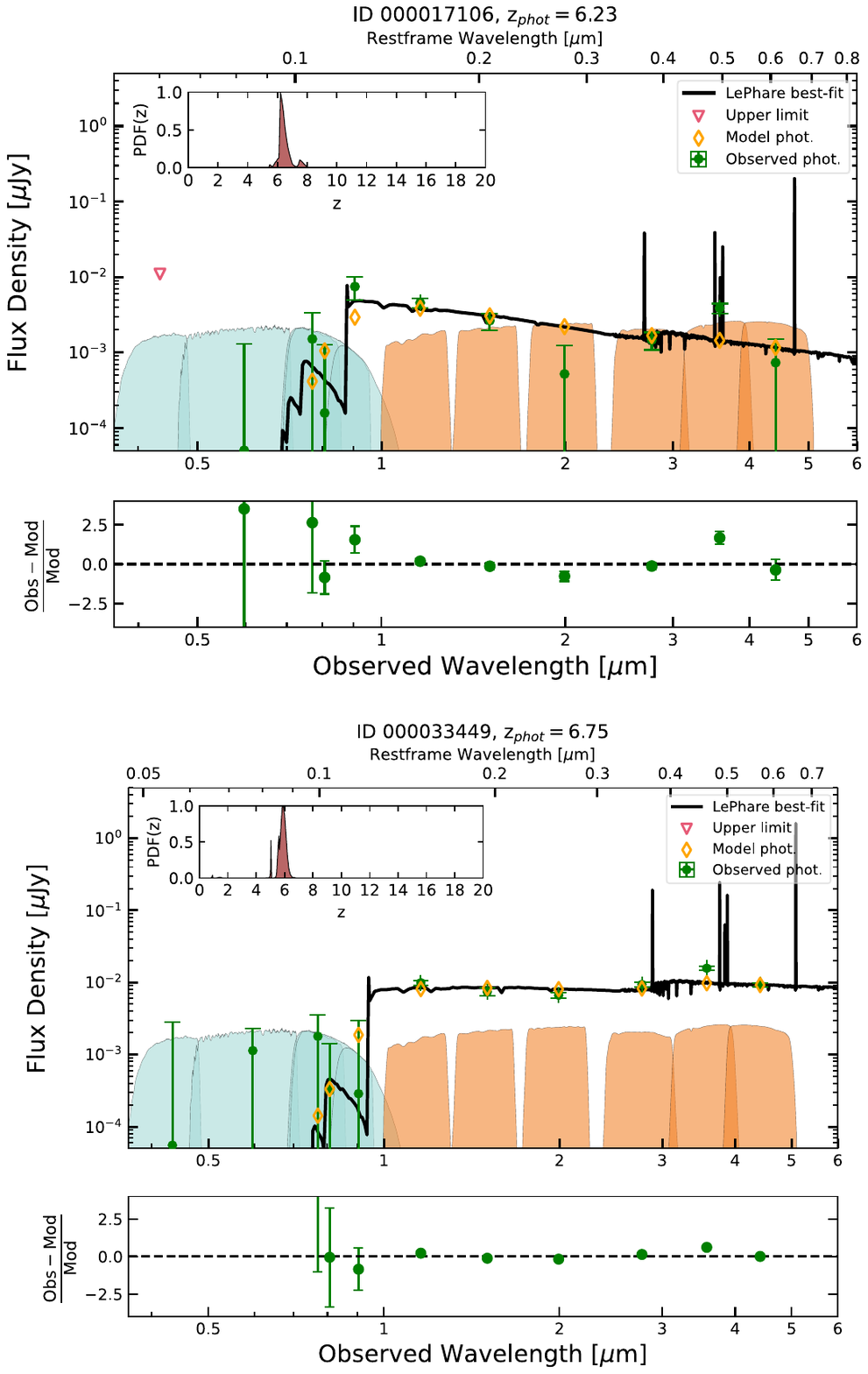}
\caption{Examples of best-fit SEDs for two of our \hbo emitters. The flux excess with respect to the continuum produced by the \hbo line complex can be seen in the F356W filter. The green/orange shaded areas show the \textit{HST/JWST} filter transmission curves. In both the top and bottom panels there is an inset showing the probability of the photometric redshift solution PDF~($z$).} \label{fig:seds}
}
\end{figure}

\section{Selection of Strong \hbo Emitters}\label{subsec:sys}

From our galaxy catalogue in the P2/NIRCam region, we considered all the galaxies with best-fit $5.5< z_{phot}<7$, as given in the output file from \texttt{LePHARE}. To select which of these sources are prominent \hbo emitters, we followed the same technique described in \citet{Rinaldi_2023}. Briefly, we analysed which of the $5.5< z_{phot}<7$ galaxies show a flux density excess (with respect to the continuum) in the NIRCam F356W band.  To obtain the corresponding \hbo line complex rest-frame equivalent width $\rm EW_0$, we measured the difference between the observed F356W flux density and the flux density of the continuum best-fit model in the same band. The latter has been obtained by repeating the SED fitting with the fixed, previously obtained redshift and excluding the F356W filter (to prevent any bias produced by the line emission).  

We then adopted the formula provided by \citet{Marmol_2016}  to convert the flux excess into the line complex rest equivalent width, which for our case is:

\begin{equation}
\rm EW_0 = W_{F356W} \, (10^{(-0.4 \Delta mag)}-1) / (1+z),
\end{equation}

\noindent where $\rm W_{F356W}$ is the rectangular width of the F356W filter, and $\rm \Delta mag$ is the difference between the observed magnitude and the synthetic continuum magnitude in the F356W filter. To guarantee that the flux density excess in the F356W band was meaningful, we imposed that $\rm \Delta mag < - 0.1$, implying that we are sensitive to selecting galaxies with a minimum $\rm EW_0 \approx 100 \, \rm \AA$  (more precisely, $\rm EW_0 \approx 107 \, \rm \AA$ at $z=5.5$ and $\rm EW_0 \approx 87 \, \rm \AA$ at $z=7$). We obtained the error bars of each source's  $\rm EW_0$  by doing 1000 random realizations of the F356W photometry, assuming a Gaussian distribution whose r.m.s. is given by the F356W photometric errors.

To ensure that the continuum was well described by the best-fit SED, we require that $\rm \Delta mag(F277W) < 2 \times error\_mag(F2777W)$. We did not impose a similar criterion for F444W, which is the filter next to the red of F356W, because F444W may be affected by $\rm H\alpha$ emission at $5.5 \lsim  z_{phot}\lsim 7$ and because we do not have F444W coverage in some cases.

Following these criteria, we found that 34 galaxies are prominent \hbo emitters, amongst a total of  102 galaxies at  $5.5< z_{phot}<7$  in our P2/NIRCam galaxy sample. For all the emitters, we verified the flux density excess in F356W by visual inspection.  These strong \hbo emitters constitute $\simeq 33\%$ of all galaxies at $5.5< z_{phot}<7$. This percentage is very similar to that reported by \citet{Rinaldi_2023} at $z=7-8$. Our derived $\rm EW_0$ values range between $\simeq 94 \, \rm \AA$ and $1695 \, \rm \AA$, with a median of $363^{+335}_{-227} \, \rm \AA$. This is broadly consistent with the results from \citet{Endsley_2023b}, but significantly lower than the median value found by \citet{Rinaldi_2023}, i.e., $943 ^{+737}_{-194} \, \rm \AA$  at $z=7-8$. This may be suggesting a redshift evolution in the median $\rm EW_0$ values (although see comments in Section~\ref{sec:gener}). A few examples of best-fit SEDs for our \hbo emitters are shown in Fig.~\ref{fig:seds}.

 We investigated the impact of adopting different photometric excess thresholds in F356W to select the line emitters in our sample. For example, if we consider a much stricter $\rm \Delta mag(F356W) < - 0.2$~mag cut, only 26 galaxies are classified as prominent \hbo emitters. Instead, if we change the limit to  $\rm \Delta mag(F356W) < - 0.07$~mag,  we will have 38 galaxies classified as strong emitters. In any case, all the analysis that we present below is based jointly on our own data points and data points from the literature, so reasonably changing the flux excess threshold (and, thus, the selected sample of strong emitters) has little impact on our results and does not change any of our conclusions.

\section{Properties of the \hbo Emitters at $5.5 < \lowercase{z} < 8$} \label{hboprop}

\subsection{General Properties}
\label{sec:gener}

First we investigate how the $\rm EW_0$\hbo values are related to some basic galaxy properties derived from the SED fitting.

\begin{figure}[h!]
\center{
\includegraphics[width=1.0\linewidth, keepaspectratio]{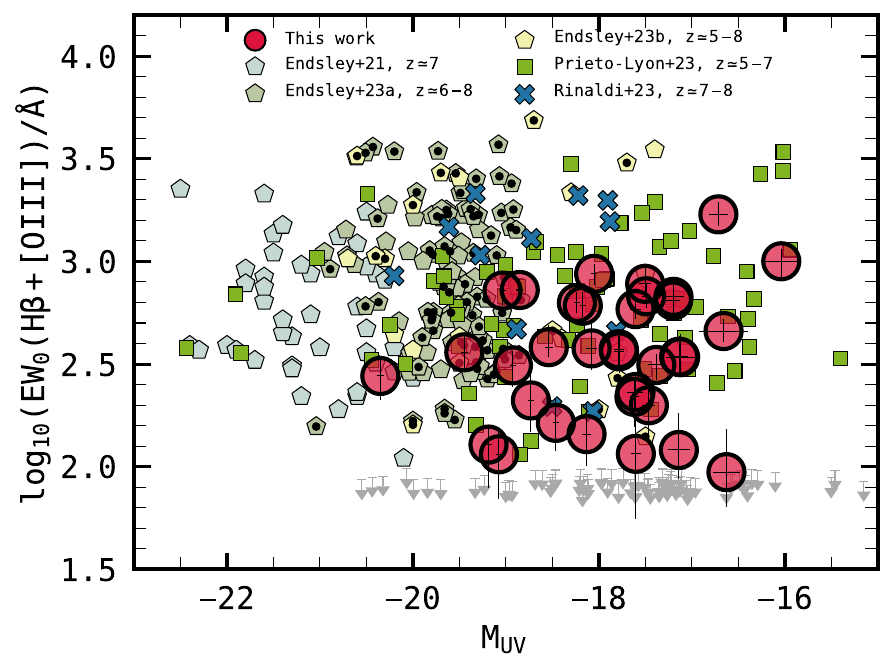}
\caption{$\rm EW_0$\hbo versus rest-UV absolute magnitude. No correlation is seen between these two parameters when all data points from this work and the literature are considered together.  The literature data points with a black dot within correspond to galaxies at $z<7$, where the dot has been added to differentiate the lower and higher redshift galaxies from a same literature sample. The gray downward-pointing arrows indicate upper limits corresponding to all the galaxies in our sample at $z=5.5-7$ which are not classified as \hbo emitters (referred to as `non-emitters' here).} \label{fig:muvew0}
}
\end{figure}

Fig.~\ref{fig:muvew0} shows the \hbo $\rm EW_0$ versus rest-UV absolute magnitude ($\rm M_{UV} = M (1500 \, \rm \AA)$) for each galaxy. In this figure, as well as all subsequent figures, we complement our data points from the P2/NIRCam \hbo line emitter sample with  \hbo emitters at $z\sim 5-8$ from the recent literature \citep{Endsley_2021, Endsley_2023a, Endsley_2023b, Prieto_Lyon_2023, Rinaldi_2023}. Most of these complementary samples also correspond to photometric selections of \hbo emitters. 

We observe no correlation between \hbo $\rm EW_0$  and $\rm M_{UV}$, which indicates that the physical processes behind these two parameters are independent. Note that $\rm M_{UV}$ does not include any dust extinction correction, as it is typically the case for $\rm M_{UV}$ in the literature. In any case, in our sample at $z=5.5-7$ only 4 out of our 34 \hbo line emitters have a best-fit colour excess $E(B-V)>0$.

\begin{figure}[h!]
\center{
\includegraphics[width=1.0\linewidth, keepaspectratio]{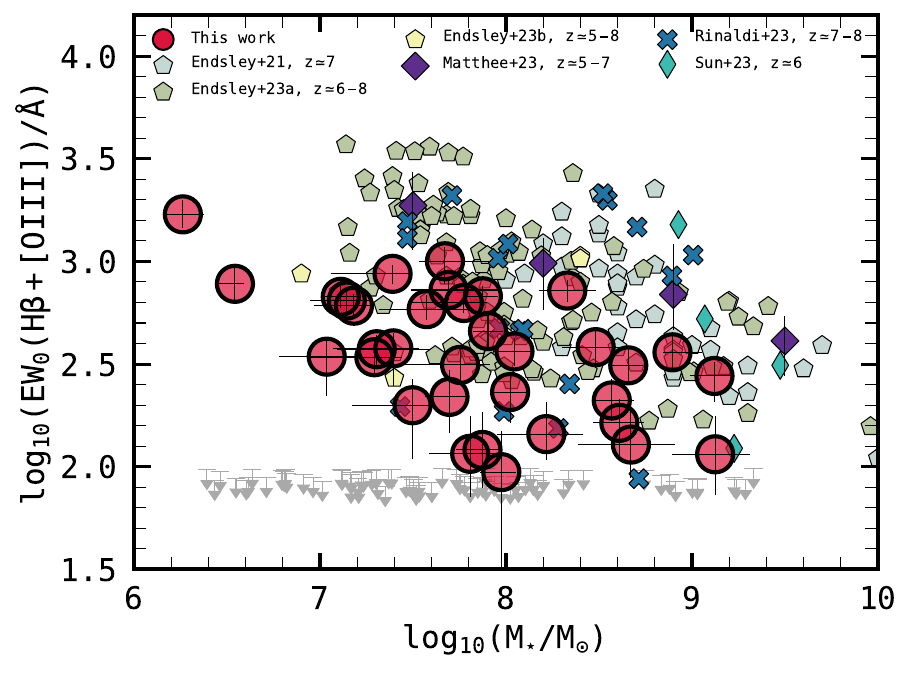}
\caption{$\rm EW_0$\hbo versus stellar mass. A broad anti-correlation between these two parameters is observed, in agreement with previous works.} \label{fig:massew0}
}
\end{figure}

With respect to the stellar mass, instead, both for our data points alone and for all data points considered together we observe a broad anti-correlation,  such that lower stellar-mass line emitters have on average higher values of  $\rm EW_0$\hbo (see Fig.~\ref{fig:massew0}). The existence of such anti-correlation was already reported at different redshifts \citep[e.g.,][]{Reddy_2018, Endsley_2021, Rinaldi_2023}. 

Our galaxy sample reaches stellar masses about 1~dex lower ($\rm log_{10}(M^\star) = 6-7$) than any other from the samples shown in Fig.~\ref{fig:massew0}, and lower than the minimum stellar masses  probed at those redshifts with other \textit{JWST} galaxy surveys \citep[e.g., ][]{Navarro_2023}. However, our galaxy sample is still incomplete at $\rm log_{10}(M^\star)< 7$.  At these lowest stellar masses, we find only two prominent \hbo emitters, both with $\rm EW_0$\hbo$\gsim 500 \, \rm \AA$, but we cannot discard the existence of other more modest line emitters with similarly low stellar masses. In addition to these two emitters with $\rm log_{10}(M^\star) = 6-7$, our galaxy sample at $z=5.5-7$ in P2/NIRCam contains 13 galaxies in the same stellar mass range which show no significant \hbo flux density excess in the F356W filter (i.e., $\rm EW_0$\hbo$\lsim 100 \, \rm \AA$; upper limits in Fig.~\ref{fig:massew0}).

An important effect observed in Fig.~\ref{fig:massew0} is the lack of galaxies with stellar mass $\rm log_{10}(M^\star) > 9$ and  $\rm EW_0$\hbo$\gsim 700 \, \rm \AA$ (Fig.~\ref{fig:massew0}). All the datasets considered here are deep enough to be basically complete at such stellar masses. Galaxies with $\rm log_{10}(M^\star) > 9$ and  emission lines with $\rm EW_0 \gsim 700 \, \rm \AA$ do exist at high redshifts \citep[e.g., ][]{Smit_2016,Caputi_2017}, but are rare and can typically be found only in large-area surveys. In the small-area surveys that we consider here, 
 line emitters with such high $\rm EW_0$\hbo are only found amongst galaxies of stellar masses $M^\star < 10^9 \, \rm M_\odot$.  We argue below that the observed trend between   $\rm EW_0$\hbo and  stellar mass is mainly produced by the dependence of $\rm EW_0$\hbo with galaxy age.

\begin{figure}[h!]
\center{
\includegraphics[width=1.0\linewidth, keepaspectratio]{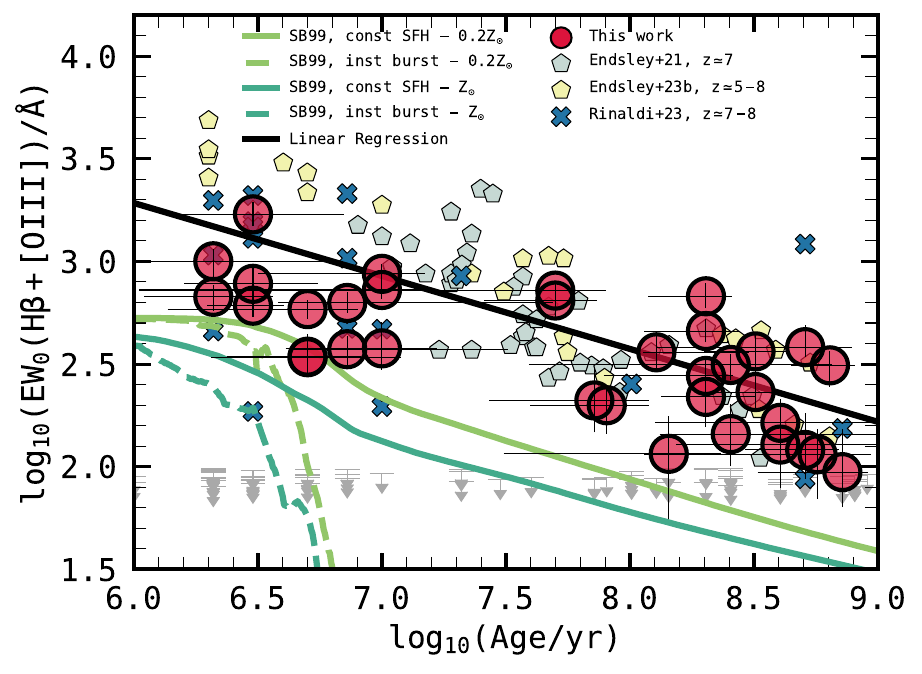}
\caption{$\rm EW_0$\hbo versus galaxy age. We also see an anti-correlation between these two parameters, which is related to the anti-correlation observed with respect to stellar mass (cf. Fig.~\ref{fig:massew0}).  The solid black line corresponds to the best-fit linear regression performed on our data points along with the data from the literature.  The model tracks from \texttt{STARBURST99} shown with green lines correspond only to the H$\beta$ $\rm EW_0$ and, thus, lie all below the observed data points.} \label{fig:ageew0}
}
\end{figure}

Fig.~\ref{fig:ageew0} shows the $\rm EW_0$\hbo versus the galaxy best-fit age. In this plot we see a similar effect to that in Fig.~\ref{fig:massew0}: there is a broad anti-correlation which is valid for the identified strong line emitters.  We performed a linear regression taking into account all data points and their errors.  We obtained a slope $\alpha=-0.36^{+0.02}_{-0.01}$. About 35\% of the emitters have very young ages ($\lsim 30 \, \rm Myr$), while the  remaining 65\% correspond to older galaxies.

The youngest galaxies (with ages $\lsim 30 \, \rm Myr$) display a relatively wide range of $\rm EW_0$\hbo values ($\rm EW_0$\hbo$\gsim 300 \, \rm \AA$).  Such a result can be explained with synthetic galaxy models and is related to the galaxy star formation history,  as can be seen in the stellar tracks in Fig.~\ref{fig:ageew0}: a galaxy passing through an instantaneous star-formation burst will suffer a quick decline of the emission line $\rm EW_0$ in only $\sim 10 \, \rm Myr$.   Being young is a necessary (albeit not sufficient) condition for a galaxy to be amongst the strongest line emitters.

At the same time, virtually no galaxy with age  $\gsim 30 \, \rm Myr$ has  $\rm EW_0 \gsim 700 \, \rm \AA$, even if they are still relatively strong emitters with $\rm EW_0$\hbo of up to several hundred $\rm \AA$. At these older ages, the line emission likely indicates more extended star formation histories or possibly an early rejuvenation effect, as those that are more common at lower redshifts \citep[e.g.,][]{Rosani_2020, Iani_2023}.

Finally, we note that there is a large percentage ($\simeq 67\%$) of galaxies in our sample at $z=5.5-7$  which do not show any F356W flux density excess, i.e., they have \hbo~$\lsim 100 \, \rm \AA$,  and span all ages between $10^6$ and $10^9 \, \rm yr$.

\subsection{Dependence on SFR}
\label{sec:sfrdep}

In this Section we investigate how the galaxy $\rm EW_0$\hbo is related to its SFR and position on the SFR-M$^\star$ plane.

\begin{figure}[h!]
\center{
\includegraphics[width=1.0\linewidth, keepaspectratio]{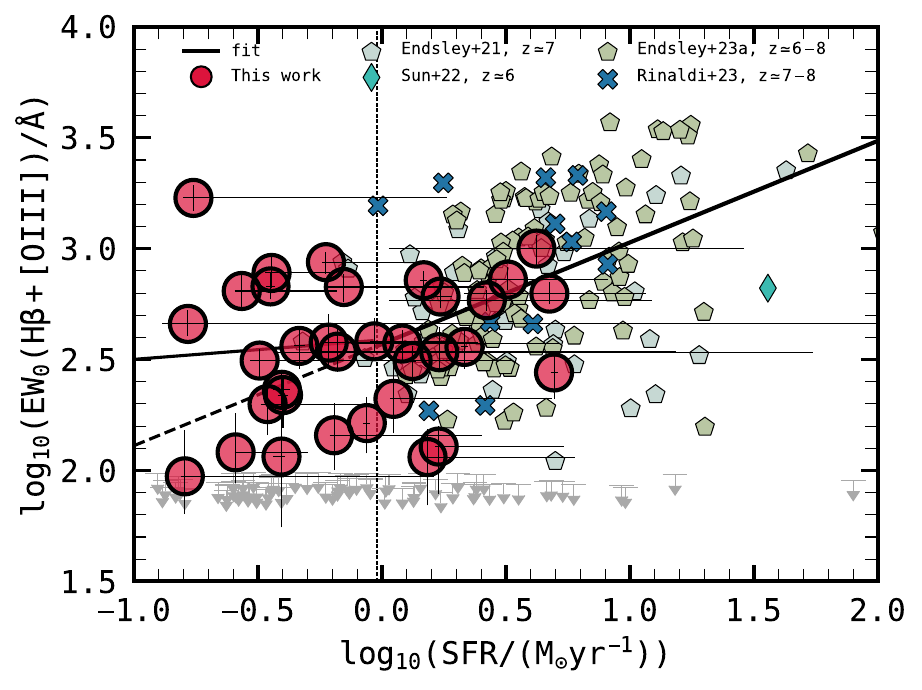}
\caption{$\rm EW_0$\hbo versus galaxy SFR. Symbols are the same as in previous figures.  The solid lines show the result of a two-component linear regression, which indicates a break in the $\rm EW_0$-SFR relation at $\rm log_{10}(SFR/M_\odot yr^{-1}) = -0.02^{+0.08}_{-0.05}$. The dashed line indicates the extrapolation of the higher SFR component, which helps to show how the relation flattens out at low SFR values.  The correlation break is produced by the high dispersion observed in the low-SFR regime, with some very high $\rm EW_0$\hbo  values. This is likely the consequence of the dominant effect of [OIII], and thus very low metallicities, in some galaxies at lower SFR.} \label{fig:sfrew0}
}
\end{figure}

Fig.~\ref{fig:sfrew0} shows the relation between $\rm EW_0$\hbo and SFR for our galaxies as well as other sources from the literature. The SFR has been derived from the rest-UV ($\lambda_{\rm rest} = 1500 \, \rm \AA$) galaxy luminosity in each case, so it is independent of the $\rm EW_0$\hbo measurement.  In turn, this UV luminosity has been obtained from the observed photometry in the filter that most closely encompasses the galaxy rest-frame $1500 \, \rm \AA$  light at the source redshift. Thanks to the depth of the NIRCam imaging in P2/NIRCam, we can probe galaxies down to unprecedented low SFR values, i.e. SFR~$\approx 0.1 \, \rm M_\odot/yr$, at $z=5.5-7$.

In contrast to the quite monotonic trend observed with respect to other galaxy parameters (age, stellar mass), the  relation between $\rm EW_0$\hbo and SFR  shows a more complex behaviour. Considering our data points jointly with those from the literature, we see that $\rm EW_0$\hbo and SFR broadly correlate with each other at SFR~$\gsim 1 \, \rm M_\odot/yr$, but this correlation flattens out at smaller SFR values.  We performed a two-component Bayesian linear regression taking into account all data points and their errors, using the \texttt{python} tool \texttt{pyro} \footnote{https://pyro.ai/examples/bayesian\_regression.html}. We obtained that the break point of the $\rm EW_0$-SFR relation occurs at $\rm log_{10}(SFR/M_\odot yr^{-1})= -0.02^{+0.08}_{-0.05}$. The slope of the relation changes from $0.46\pm0.02$ at higher SFRs to $0.09^{+0.09}_{-0.08}$  at lower SFRs.
Indeed, at  SFR~$\lsim 1 \, \rm M_\odot/yr$, galaxies display a wide range of possible $\rm EW_0$\hbo values and no correlation is observed any more with the galaxy SFR.

In principle, one would expect that the behaviour of $\rm EW_0$\hbo versus SFR  is similar to  that of $\rm EW_0$\hbo versus $\rm M_{UV}$. However, the $\rm M_{UV}$ values in Fig.~\ref{fig:muvew0} are not corrected for dust extinction (as it is usual in the literature), while the SFR values are. This only affects four (out of 34) emitters in our sample, but also brighter galaxies from the other considered datasets.  The four emitters whose SFR are dust-corrected end up having  SFR~$ > 1 \, \rm M_\odot/yr$, so they make part of the positive correlation observed at these higher SFR values, but they are not responsible for it.

The positive correlation  between $\rm EW_0$\hbo and SFR obtained at SFR~$ > 1 \, \rm M_\odot/yr$ is mainly driven by the data points from the literature. Note also that, for some of them, the SFR values have been calculated from best-fit SED models, in contrast to our own values and those of, e.g., \citet{Rinaldi_2023},  which have been empirically obtained directly from the galaxy rest-UV luminosities and are independent of the SED fitting. In any case, a comparison of these methodologies to compute the SFR (based on our own data) indicates that there should not be any systematic effect and, therefore, the resulting break  observed in the correlation at SFR~$ \simeq 1 \, \rm M_\odot/yr$ in Fig.~\ref{fig:sfrew0} should be robust against these methodology differences.

 We tested the sensitivity of the $\rm EW_0$-SFR correlation break to our adopted threshold of flux density excess to select strong line emitters.  As explained above, if we lowered the threshold to  $\rm \Delta mag(F356W) < - 0.07$~mag, then there would be 38 strong \hbo emitters in our sample (instead of 34). We repeated the two-component linear regression considering all these galaxies along with the literature data points. We found that the break is similarly significant  as that show in Fig.~\ref{fig:sfrew0} and the break point shifts only slightly to  $\rm log_{10}(SFR/M_\odot yr^{-1}) \approx -0.05$.  We also checked that the break is not driven by the inclusion of literature data points up to $z=8$,  i.e. beyond the redshift limits of our own sample. If we repeat the analysis considering only literature data points up to $z=7$, we still find a significant correlation break with a break point at $\rm log_{10}(SFR/M_\odot yr^{-1}) \approx -0.08$.

This change of trend in the \hbo line complex behaviour suggests that the  \hbo line complex might be dominated by different physical processes at low and high star formation rates. The positive correlation at SFR~$\gsim 1 \, \rm M_\odot/yr$  indicates that the strength of [OIII] must be following the SFR, as H$\beta$ more obviously does. At lower SFR, instead, H$\beta$  must become less important and, thus,  the $\rm EW_0$\hbo value might be virtually dominated by [OIII]. The high dispersion observed in this low-SFR regime, with some very high $\rm EW_0$\hbo  values,  suggests that decreasing gas metallicities may be the main reason for the increasing $\rm EW_0$\hbo with decreasing SFR.  Low metallicities are linked to higher radiation fields \citep[e.g., ][]{Kumari_2021}, which in turn produce more prominent nebular lines, i.e., nebular lines with higher equivalent widths.

A roughly similar effect is observed from spectroscopic studies at lower redshifts, particularly Fig.~6 in \citet{Reddy_2018}, although in this work the SFR regimes below and above SFR~$\approx 1 \, \rm M_\odot/yr$ are probed at different redshifts ($z=0$ and $z\sim 1.5-3.8$, respectively). The separate analyses of the H$\beta$ and [OIII] EW in that work indicate that, as expected,  it is the [OIII] emission line the one that drives the trend break at low SFR.

As it was the case for low stellar-mass galaxies, many of the galaxies with  SFR~$< 1 \, \rm M_\odot/yr$ in our sample are actually not prominent line emitters, i.e. have $\rm EW_0$\hbo$ \lsim 100 \, \rm \AA$  (shown as upper limits in Fig.~\ref{fig:sfrew0}). So once again the discussed trends only apply to the subset of galaxies that do show line emission. As we discuss throughout this paper, if star formation activity proceeds in bursts rather than continuously, then emission lines are only expected to be present at very young ages ($\lsim 10 \, \rm Myr$). Instead, the rest-UV continuum emission indicative of ongoing star formation lasts longer.  

As a matter of fact, at young galaxy ages, deriving SFR from rest-UV fluxes in the canonical manner is not strictly correct. Most SFR tracers stabilize only after $\sim 100 \, \rm Myr$ \citep{OtiFloranes_2010} and indeed the \citet{Kennicutt_1998} prescription assumes a constant star formation history for $100 \, \rm Myr$. The youngest galaxies in our sample, including most of the line emitters, do not comply with this assumption.  We tested the impact of deriving SFR values based on rest-UV luminosities with the \citet{Kennicutt_1998} empirical law for these very young sources, in order to understand whether their presence has any influence in our conclusions. For this, we corrected the derived SFR by the expected SFR(H$\alpha$)/SFR(UV) ratios at different ages, following the tracks shown in Fig.~15 of \citet{Iani_2023}, which are in turn based on BPASS synthetic galaxy models \citep{Eldridge_2017, Stanway_2018}.  We found that applying these corrections does not have any significant impact in our results and conclusions.

\subsection{The SFR-M$^{\star}$ plane}

\begin{figure}[h!]
\center{
\includegraphics[width=1.0\linewidth, keepaspectratio]{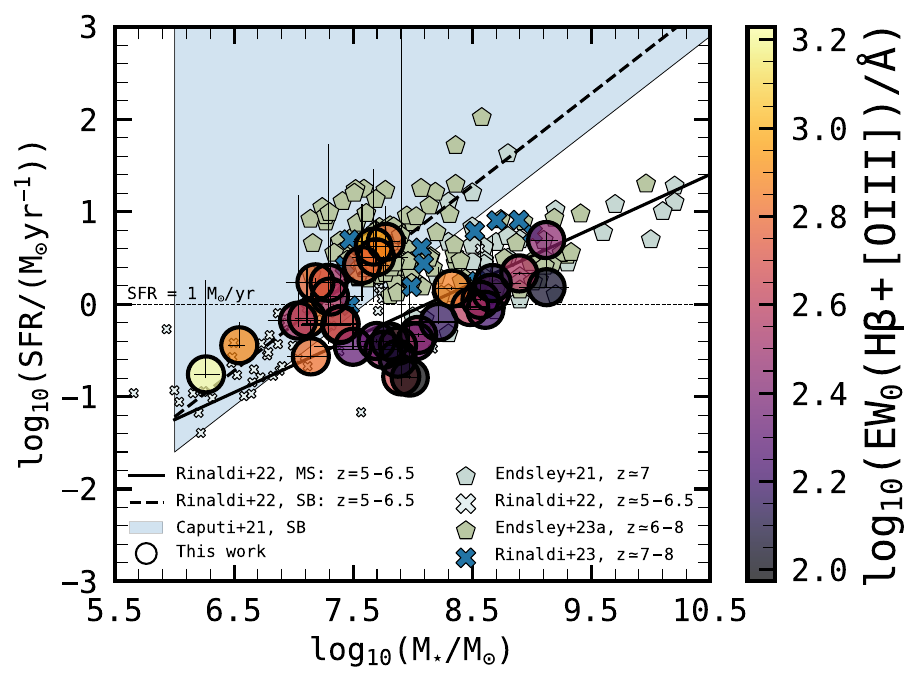}
\caption{Location of our galaxies in the SFR-M$^\star$ plane, colour-coded according to their $\rm EW_0$\hbo values. As in previous figures, we complement our galaxy sample with other samples from the literature at $z=5-8$ \citep{Endsley_2021, Rinaldi_2022, Endsley_2023a, Prieto_Lyon_2023, Rinaldi_2023}. The light-blue shaded region indicates the starburst zone, as defined in \citet{Caputi_2017, Caputi_2021}. } \label{fig:sfrstm}
}
\end{figure}

Fig.~\ref{fig:sfrstm} shows the location of our \hbo emitters at $z=5.5-7$, along with other galaxies at $z=5-8$ from the literature, on the SFR-M$^{\star}$ plane. Our \hbo emitters are colour-coded according to their $\rm EW_0$\hbo values. 

The \hbo emitters occupy different regions of the SFR-M$^\star$ plane, with some being located in the star-formation main sequence \citep[e.g.,][]{Speagle_2014} and others in the starburst zone, empirically defined as the half-plane with $\rm log_{10}(sSFR(yr^{-1}))>-7.6$ \citep{Caputi_2017, Caputi_2021}. According to this definition, starbursts are galaxies with a stellar-mass doubling time $\lsim 40 \rm \, Myr$, consistently with local starbursts \citep{Kennicutt_1998}. From  Fig.~\ref{fig:sfrstm}, it is evident that the strongest line emitters are preferentially found amongst the starburst galaxies. These \hbo emitters in the starburst zone span a wide range of stellar masses, between $\rm log_{10}(M^\star/M_\odot) \approx 6$  and $\approx 9$.

We see that the star-formation main sequence and starburst cloud  converge on the SFR-M$^{\star}$ plane at SFR~$\approx 0.3 \, \rm M_\odot/yr$, i.e., a somewhat lower value than the break SFR discussed in \S\ref{sec:sfrdep}. This intersection was predicted by \citet{Rinaldi_2022} from the extrapolation of the two star-forming mode trends towards low stellar masses at high redshifts. The depth of our new\textit{JWST} data in P2/NIRCam allows us to directly detect a few galaxies there. In this regime of low stellar masses and star formation rates, all star formation should be proceeding in a single mode and the gap between the star-formation main sequence and starbursts disappears.

\begin{figure}[h!]
\center{
\includegraphics[width=1.0\linewidth, keepaspectratio]{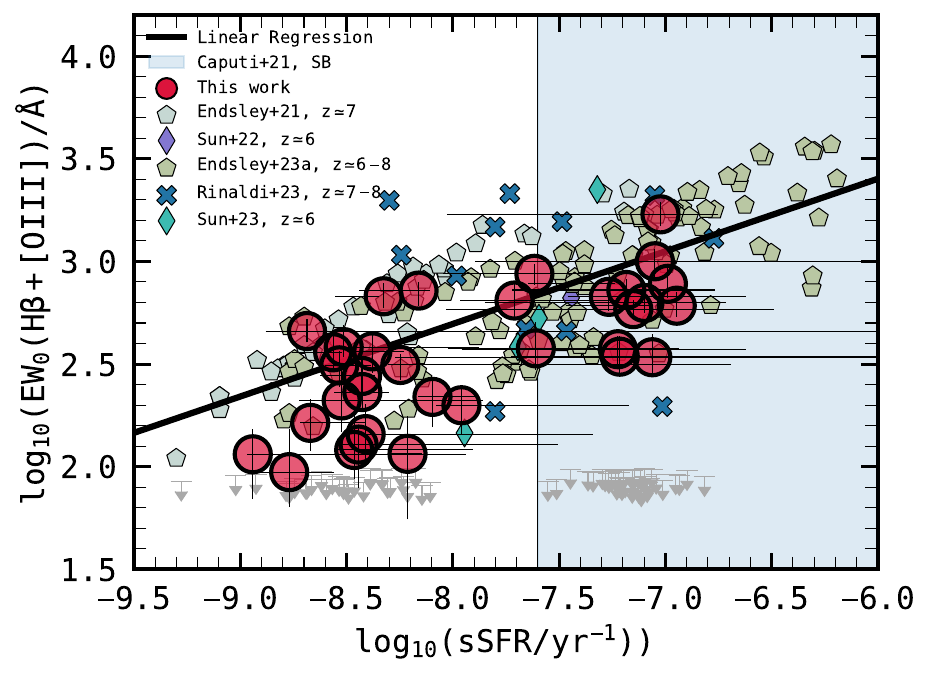}
\caption{$\rm EW_0$\hbo versus galaxy sSFR. symbols are the same as in previous figures. The light-blue shaded region indicates the starburst zone, as defined in \citet{Caputi_2017, Caputi_2021}.} \label{fig:ew0ssfr}
}
\end{figure}

Finally, we show the relation between $\rm EW_0$\hbo and sSFR  in Fig.~\ref{fig:ew0ssfr}. Here we also observe a positive correlation, such that the most prominent \hbo emitters tend to have higher sSFR values. In this case, for the best-fit linear regression we obtain a slope $\alpha=0.35\pm0.01$. Particularly, the most extreme emitters are mostly found in the starburst zone. For example, if we consider all those galaxies  with $\rm EW_0 > 700 \, \rm \AA$, we find that $\simeq 70\%$ of them are starbursts. Instead, among the line emitters with lower $\rm EW_0$\hbo, only $\simeq 25\%$ are in the starburst region.

A cross-correlation between $\rm EW_0$\hbo and sSFR, even if broad, is not trivial. We do expect a galaxy sSFR to be more directly related to the $\rm EW_0 (H\alpha)$, as the $\rm H\alpha$ luminosity provides a fiducial measurement of the galaxy SFR, while the continuum at the $\rm H\alpha$ rest-frame wavelength is roughly proportional to the galaxy stellar mass. Instead, the \hbo line complex is expected to be dominated by the [OIII] emission in most cases \citep[e.g.,][]{Cameron_2023, Langeroodi_2023}, making the total \hbo luminosity to depend not only on the galaxy SFR, but also, e.g., its gas temperature and metallicity. All these properties will affect the $\rm EW_0$\hbo  values.

In any case, it is important to note that the observed trend between  $\rm EW_0$\hbo and sSFR  also  applies exclusively to galaxies with identified \hbo emission. In our own sample in P2/NIRCam, we have 68 galaxies at $z=5.5-7$ with no \hbo flux density excess in the F356W filter.  And, amongst these non-emitters, $\simeq 55\%$ are starburst galaxies, although this percentage should be considered an upper limit because a flux-limited galaxy survey will mainly prevent the selection of galaxies with low sSFR at fixed stellar mass (Rinaldi et al., in prep.). In summary, the starburst region does not contain only galaxies with high $\rm EW_0$\hbo values, but those which do have such high $\rm EW_0$ values are preferentially found in a starbursting phase.

As discussed before, the fact that there are many galaxies which are starbursts, but have low values of $\rm EW_0$\hbo is likely related to the fact that the rest-UV galaxy luminosity and the Balmer lines do not trace star formation activity over the same timescales \citep[e.g., ][]{Sparre_2017,Faisst_2019}. The rest-UV luminosity is produced by O and B-type stars, so it typically indicates star formation activity on timescales of  $\sim 100 \, \rm Myr$. Instead,  in galaxies passing through a burst of star formation, Balmer lines will sharply decrease their luminosities after a few $\times 10 \, \rm Myr$ \citep[e.g.,][]{IglesiasParamo_2004, Lee_2009,OtiFloranes_2010, Guo_2016, Emami_2019, Iani_2023}.

\section{Summary and Discussion} \label{sec:conc}

In this paper we investigated the dependence of strong \hbo emission ($\rm EW_0$\hbo~$\gsim 100 \, \rm \AA$) on the main host galaxy properties, particularly those that are derived from SED fitting, at $z=5.5-8$.  Considering jointly our own data and recent results from the literature has been important to increase the statistics of our analysis, as well as (at least partly) homogenizing the possible selection effects from the different datasets. Moreover, by studying also the non-emitters in our sample at comparable redshifts, we could also put the emitters in the more general context of galaxy evolution around the Epoch of Reionization.

For the strong \hbo emitters we found broad anti-correlations between the $\rm EW_0$\hbo and both galaxy stellar mass and age. These two $\rm EW_0$\hbo  anti-correlations are not independent: the most massive galaxies ($\rm log_{10}(M^\star/M_\odot) > 8$)  are amongst the oldest ones ($\rm log_{10} (age)\gsim 7.5$) at $z=5-8$. However, the roles of these two parameters are different. While a higher stellar mass would mainly decrease the $\rm EW_0$\hbo by increasing the underlying continuum light, galaxy age would directly affect the emission line luminosities, as it is expected from galaxy spectral models \citep{SB99}. Similar trends have previously been reported in the literature at different redshifts \citep[e.g.,][]{Khostovan_2016, Reddy_2018,  Endsley_2021, Boyett_2022, Matthee_2023}, which suggests that the physics driving strong line emitters is basically the same through cosmic time.

We observe only a very tentative evolution of these trends in the redshift range analysed here $z=5.5-8$. This is perhaps not surprising given the corresponding short elapsed time ($\sim 0.5 \, \rm Gyr$), but still interesting to remark given that this period comprises the epoch before and after the end of Reionization.

A key result of this paper is the finding that the relation between  $\rm EW_0$\hbo  and the galaxy SFR could change at $\rm SFR \lsim 1 \, \rm M_\odot/yr$. The unprecedented depth of our NIRCam data allows us to explore  such low SFR values for non-lensed galaxies at $z=5.5-7$.  Indeed, the correlation observed between $\rm EW_0$\hbo and SFR at $\rm SFR \gsim 1 \, \rm M_\odot/yr$ flattens out at $\rm SFR \lsim 1 \, \rm M_\odot/yr$, which sugggests that in this regime the \hbo complex is dominated by the [OIII] line and galaxies may have lower metallicities towards lower SFR values. Indeed, this has been shown to be the case at low redshifts \citep{Duarte_2022}. Low metallicities are linked to high radiation fields, which in turn are responsible for higher luminosities (and equivalent widths) in nebular lines. Unfortunately the current data do not allow us to directy constrain the galaxy metallicities, except crudely via the SED modelling. As expected, the majority ($\gsim 70\%$) of our galaxies have a best-fit SED with sub-solar metallicity, including those with the lowest stellar-mass and SFR values. 

We note that incompleteness may be affecting our sample at $\rm SFR \lsim 1 \, \rm M_\odot/yr$, i.e. there may be undetected galaxies which have such low SFRs and significant \hbo emission (with $\rm EW_0$\hbo~$\gsim 100 \, \rm \AA$).  These possibly missing galaxies could lie close to the extrapolation of the $\rm EW_0$\hbo - SFR correlation for $\rm SFR \gsim 1 \, \rm M_\odot/yr$ at $\rm SFR \lsim 1 \, \rm M_\odot/yr$, but still the sources with low SFR and high $\rm EW_0$\hbo will be present, indicating that a simple correlation cannot explain the behaviour of all sources at low SFR.

Another possibility to explain the break in the $\rm EW_0$\hbo versus SFR relation at $\rm SFR \approx 1 \, \rm M_\odot/yr$ is that at low SFR the spectral line emission can be affected by stochastic sampling of the galaxy star formation and/or IMF \citep[e.g.,][]{Boissier_2007, Lee_2009, daSilva_2012, Forero_2013, MasRibas_2016}. These kinds of effects only matter at very low SFR ($\sim 0.1 \, \rm M_\odot/yr$) at low redshifts, but the bursty nature characterising many high-$z$ galaxies may still produce a stochastic sampling of the star-forming units within a galaxy \citep{Vikaeus_2020, Pallottini_2023}.  This, in turn, can have important implications for the estimation of metallicities in low-SFR galaxies at high redshifts \citep[e.g.,][]{Vanzella_2023}.

In any case, we note that our results suggesting low metallicities for starbursting low stellar-mass galaxies are in line with the predictions of galaxy formation models. Using the \texttt{ASTRAEUS} galaxy formation framework \citep{Hutter_2021}, which couples galaxy formation with Reionization,  \citet{Ucci_2023} showed that the mass-metallicity relation of galaxies around the Epoch of Reionization depends on their sSFR, such that higher sSFR values correspond to lower metallicities at fixed stellar mass. This is explained because galaxies with higher sSFR had stronger outflows and, thus, a higher amount of metal ejection, leaving their interstellar medium less metal-enriched.

The location of the \hbo emitters on the SFR-M$^\star$ plane shows also a break in the star-formation MS/starburst bimodality towards low SFR values, albeit at somewhat lower SFR  than the SFR turning point in the  $\rm EW_0$\hbo-SFR relation. The convergence of the two sequences was discussed by \citet{Rinaldi_2022} and suggests that all star formation happens in a single mode at the lowest stellar-mass galaxies. Our current ultra-deep observations allow us to directly see a few galaxies in this convergence regime.

In general, the highest $\rm EW_0$\hbo values are also associated with high sSFR and, correspondingly, with a high incidence of the strongest \hbo emitters in the starburst region of the SFR-$\rm M^\star$ plane ($\sim 70\%$ of those with $\rm EW_0$\hbo~$>700 \, \rm \AA$ are starbursts). This result agrees with the findings of \citet{Boyett_2022}  and \citet{Endsley_2023a}, who suggested that the brightest line emitters at high redshifts could be experiencing a strong upturn in their SFR. It is also in line with the results derived from the FirstLight galaxy simulation \citep{Ceverino_2017,Ceverino_2018}, which predict a correlation between $\rm EW_0$[OIII] and galaxy sSFR, although only a very small fraction of their simulated galaxies have the high sSFR values characterising our starburst galaxies at $z=5.5-7$ \citep{Ceverino_2021}.

Considering the galaxy properties analysed here altogether allow us to conclude that the strongest line emitters are typically young, low stellar-mass galaxies that are starbursting or very close to the starburst phase. These makes them favourite candidates for the sources of Reionization, as their ionizing photon production efficiency could be significantly higher than for other galaxies \citep{Izotov_2018, Rinaldi_2023b, Simmonds_2023}.

Throughout this work we have analysed the properties of the strong \hbo emitters at $z=5.5-7$ in comparison to those of all other sources (which we called `non-emitters') at similar redshifts. The strong emitters constitute only $\sim$33\% of all the galaxies at $z=5.5-7$ in our P2/NIRCam sample, in broad agreement with the percentage reported in the literature for $z=7-8$ \citep[][]{Rinaldi_2023}. We found that many non-emitters share the same properties as the typical strong emitters, i.e.,  low stellar masses, young ages and high sSFR. This strongly suggests that the strong line emitters are not different in nature to many other galaxies at $z=5.5-7.0$, they are rather the same kinds of galaxies just passing through the initial stages of a burst of star formation. At their flux limit, galaxy surveys will preferentially contain strong line emitters, as the emission line can boost their observability \citep[e.g.,][]{Sun_2023}.

These results considered together suggest that {\em all} young, low stellar-mass, star-forming galaxies at such high redshifts could have had a role in Reioization. Their importance as ionizing sources was maximum at the beginning of the starburst phase. Ultra-deep spectroscopic studies to be conducted with \textit{JWST} are necessary to better understand the physical conditions associated  with star formation in young, low stellar-mass galaxies.

\acknowledgments

In memoriam to the MIRI European Consortium members Hans-Ulrik N\o{}rgaard-Nielsen and Olivier Le F\`evre.

The authors thank an anonymous referee for a constructive report. They also thank Ryan Endsley and Jorryt Matthee for providing their galaxy sample data in electronic format, and Pratika Dayal for useful discussions. 

This work is based on observations made with the NASA/ESA/CSA James Webb Space Telescope. The data were obtained from the Mikulski Archive for Space Telescopes at the Space Telescope Science Institute, which is operated by the Association of Universities for Research in Astronomy, Inc., under NASA contract NAS 5-03127 for JWST. These observations are associated with programs GO \#2079 and GTO \#1283. The authors acknowledge the team led by Steven Finkelstein, Casey Papovich and Norbert Pirzkal for developing their respective observing programs with a zero-exclusive-access period. Also based on observations made with the NASA/ESA Hubble Space Telescope obtained from the Space Telescope Science Institute, which is operated by the Association of Universities for Research in Astronomy, Inc., under NASA contract NAS 5-26555. 
.
The work presented here is the effort of the entire MIRI team and the enthusiasm within the MIRI partnership is a significant factor in its success. MIRI draws on the scientific and technical expertise
of the following organisations: Ames Research Center, USA; Airbus Defence and Space, UK; CEA-Irfu, Saclay, France; Centre Spatial de Li{\`e}ge, Belgium; Consejo Superior de Investigaciones Cientificas, Spain; Carl Zeiss Optronics, Germany; Chalmers University of Technology, Sweden; Danish Space Research
Institute, Denmark; Dublin Institute f\"ur Advanced Studies, Ireland; European Space Agency, Netherlands; ETCA, Belgium; ETH Zurich, Switzerland; Goddard Space Flight Center, USA; Institute d'Astrophysique Spatiale, France; Instituto Nacional de T\'ecnica Aeroespacial, Spain; Institute for Astronomy, Edinburgh, UK; Jet Propulsion Laboratory, USA;  Laboratoire d'Astrophysique de Marseille (LAM), France; Leiden University, Netherlands; Lockheed Advanced Technology Center (USA); NOVA Opt-IR group at Dwingeloo, Netherlands; Northrop Grumman, USA; Max-Planck Institut f\"ur Astronomie (MPIA), Heidelberg, Germany; Laboratoire dEtudes Spatiales et d'Instrumentation en Astrophysique (LESIA), France; Paul Scherrer Institut, Switzerland; Raytheon Vision Systems, USA; RUAG Aerospace, Switzerland; Rutherford Appleton Laboratory (RAL Space), UK; Space Telescope Science Institute, USA; Toegepast-
Natuurwetenschappelijk Onderzoek (TNO-TPD), Netherlands; UK Astronomy Technology Centre, UK; University College London, UK; University of Amsterdam, Netherlands; University of Arizona, USA; University of Cardiff, UK; University of Cologne, Germany; University of Ghent; University of Groningen, Netherlands; University of Leicester, UK; University of Leuven, Belgium; University of Stockholm, Sweden; Utah State University, USA.

KIC and EI acknowledge funding from the Netherlands Research School for Astronomy (NOVA). KIC acknowledges funding from the Dutch Research Council (NWO) through the award of the Vici Grant VI.C.212.036.  PGP-G and LC acknowledge support from the Spanish Ministerio de Ciencia e Innovaci\'on MCIN/AEI/10.13039/501100011033 through grant PGC2018-093499-B-I00. LC acknowledges financial support from Comunidad de Madrid under Atracci\'on de Talento grant 2018-T2/TIC-11612.  G\"O \&  JM  acknowledge support from the Swedish National Space Administration (SNSA). TRG, SG and IJ acknowledge funding from the Cosmic Dawn Center (DAWN), funded by the Danish National Research Foundation (DNRF) under grant DNRF140. SG acknowledges financial support from the Villum Young Investigator grant 37440 and 13160.  JH and DL were supported by a VILLUM FONDEN Investigator grant to JH (project number 16599). JAM and ACG acknowledge support by grant PIB2021-127718NB-100 by the Spanish Ministry of Science and Innovation/State Agency of Research MCIN/AEI/10.13039/ 501100011033 and by ``ERDF A way of making Europe''.

%% To help institutions obtain information on the effectiveness of their 
%% telescopes the AAS Journals has created a group of keywords for telescope 
%% facilities.
%
%% Following the acknowledgments section, use the following syntax and the
%% \facility{} or \facilities{} macros to list the keywords of facilities used 
%% in the research for the paper.  Each keyword is check against the master 
%% list during copy editing.  Individual instruments can be provided in 
%% parentheses, after the keyword, but they are not verified.

\vspace{5mm}
\facilities{ALMA, VLT, \textit{HST}, \textit{Spitzer}, \textit{Herschel}}

%% Similar to \facility{}, there is the optional \software command to allow 
%% authors a place to specify which programs were used during the creation of 
%% the manusscript. Authors should list each code and include either a
%% citation or url to the code inside ()s when available.

\software{SExtractor, IRAF \citep{Tody_1986, Tody_1993}, LePhare}

\dataset[NIRCam (MIDIS) P2 DOI]{https://doi.org/10.17909/je9x-d314}

\dataset[NGDEEP P2 DOI]{https://doi.org/10.17909/v7ke-ze45}

\dataset[HST Legacy P2 DOI]{https://doi.org/10.17909/T91019}

\dataset[JWST Pipeline Calibration File DOI]{https://zenodo.org/records/7314521}

%% Appendix material should be preceded with a single \appendix command.
%% There should be a \section command for each appendix. Mark appendix
%% subsections with the same markup you use in the main body of the paper.

%% Each Appendix (indicated with \section) will be lettered A, B, C, etc.
%% The equation counter will reset when it encounters the \appendix
%% command and will number appendix equations (A1), (A2), etc. The
%% Figure and Table counter will not reset.

%% This command is needed to show the entire author+affilation list when
%% the collaboration and author truncation commands are used.  It has to
%% go at the end of the manuscript.
%\allauthors

%% Include this line if you are using the \added, \replaced, \deleted
%% commands to see a summary list of all changes at the end of the article.
%\listofchanges

\bibliography{References}{}
\bibliographystyle{aasjournal}
\end{document}